\documentclass{emulateapj}
\usepackage{graphicx}
\begin{document}

\def\refnew#1{(\ref{#1})}
\def\etal{et al.\ \rm}

\title{Magnetospheric eclipses in the double pulsar system J0737-3039.}

\author{Roman R. Rafikov\altaffilmark{1} and Peter
Goldreich\altaffilmark{1,2}} \altaffiltext{1}{IAS, Einstein Dr.,
Princeton, NJ 08540} \altaffiltext{2}{Theoretical Astrophysics,
California Institute of Technology, MS 130-33, Pasadena, CA 91125}
\email{rrr@ias.edu, pmg@ias.edu}


\begin{abstract}

We argue that eclipses of radio emission from the millisecond pulsar A
in the double pulsar system J0737-3039 are due to synchrotron
absorption by plasma in the closed field line region of the
magnetosphere of its normal pulsar companion B. Based on a plausible
geometric model, A's radio beam only illuminates B's magnetosphere for
about 10 minutes surrounding the time of eclipse. During this time it
heats particles at $r\gtrsim 10^9$ cm to relativistic energies and
enables extra plasma, beyond that needed to maintain the corotation
electric field, to be trapped by magnetic mirroring. An enhancement of
the plasma density by a factor $\sim 10^2$ is required to match the
duration and optical depth of the observed eclipses.  The extra plasma
might be supplied by a source near B through $B\gamma$ pair creation
by energetic photons produced in B's outer gap. Relativistic pairs
cool by synchrotron radiation close to where they are
born. Re-excitation of their gyrational motions by cyclotron absorption
of A's radio beam can result in their becoming trapped between
conjugate mirror points in B's magnetosphere. Because the trapping
efficiency decreases with increasing optical depth, the plasma density
enhancement saturates even under steady state illumination. The result
is an eclipse with finite, frequency dependent, optical depth. After
illumination by A's radio beam ceases, the trapped particles cool and
are lost.  The entire cycle repeats every orbital period. We speculate
that the asymmetries between eclipse ingress and egress result 
in part from the magnetosphere's evolution toward a steady state 
when illuminated by A's radio beam. We predict that A's linear
polarization will vary with both eclipse phase and B's rotational
phase.

\end{abstract}

\keywords{pulsars: general --- pulsars: individual (J0737-3039A, 
J0737-3039B) ---  stars: neutron --- radiation mechanisms: 
non-thermal --- plasmas}


\section{Introduction.}
\label{sect:intro}


The recent discovery of the binary pulsar J0737-3039 --- a millisecond
pulsar (pulsar A with a period $P_A=23$ ms) and a normal pulsar
(pulsar B with a period $P_B=2.8$ s) in a tight $2.4$ hrs orbit
(Burgay \etal 2003) --- not only provides us with unprecedented tests
of general relativity (Lyne \etal 2004) but also reveals a variety of
magnetospheric phenomena.  Among the latter are variations of pulsar
B's radio emission correlated with binary orbital phase (Lyne \etal 2004;
Ransom \etal 2004) and modulated at the spin frequency of pulsar A
(McLaughlin \etal 2004a), and periodic eclipses of pulsar A when
it passes behind pulsar B (Lyne \etal 2004; Kaspi \etal 2004; McLaughlin
\etal 2004b). It is the latter phenomenon that concerns us in this paper.

Detailed observations by the Green Bank Telescope (Kaspi \etal 2004)
established the following frequency-averaged properties of pulsar A's
eclipses: the eclipse duration is about $27$ s, which for a relative
transverse velocity of $680$ km s$^{-1}$ translates into a size of
$18,000$ km. Eclipses are significantly asymmetric with ingress taking
$3-4$ times longer than egress; pulsar A's radio beam is extinguished
more strongly post conjunction, consistent with zero flux, than before
conjunction when some flux leaks through. Analysis of the same data at
higher time resolution by McLaughlin \etal (2004b) uncovered effects
of B's rotational phase on the frequency dependence and shape of the
eclipse during ingress; its shape during egress is remarkably
independent of either B's rotational phase or radio frequency.

The spin-down luminosity of pulsar A exceeds that of pulsar B by about
$3600$, so it is plausible that B's magnetosphere is compressed by a
relativistic wind from A.  Calculations by Lyutikov (2004) demonstrate
that the ram pressure of A's wind can be balanced by B's magnetic
field pressure at a stand-off distance $r_{so}\approx (3.5-6)\times
10^9$ cm which is within B's light cylinder radius of $r_{L,B}\equiv
c/\Omega_B\approx 1.3\times 10^{10}$ cm. A crucial point is that the
size of the eclipsing region is considerably smaller than even the
compressed size of the B's magnetosphere. To fully quantify the
geometry of the eclipse, the inclination of the system must be
accurately known. Measurements of the Shapiro delay established that
the inclination is very high, $i=87^\circ\pm 3^\circ$ (Lyne \etal
2004). A refined estimate by Coles \etal (2004) based on correlation
of interstellar scintillations of both pulsars yields
$i=90.26^\circ\pm 0.13^\circ$, which corresponds to a minimum distance
of the A's radio beam with respect to B's position projected on the
plane of the sky of only $4000\pm 2000$ km. This refinement implies
that the extinction during eclipse arises {\it inside} pulsar B's
magnetosphere and that the radial extent of the eclipsing region is
about $10,000$ km.

Our goal is to evaluate the absorption of the radio beam of pulsar A
as it passes through the magnetosphere of pulsar B. We show in \S
\ref{sect:cyclo} that resonant cyclotron absorption in the
charged-separated, closed field line region would provide only a small
optical depth. However, we demonstrate in \S \ref{sect:synchro} that
absorption of radiation from pulsar A heats charged particles in B's
magnetosphere to relativistic energies. In \S \ref{sect:supply} we
describe how this enables the accumulation of additional, charge
neutral plasma in B's magnetosphere. As a result, the radio emission
of pulsar A can be significantly extinguished by synchrotron
absorption. We devote \S \ref{sect:disc} to a discussion of the
ramifications of our model.


\section{Resonant Cyclotron Absorption.}
\label{sect:cyclo}


The region of closed field lines in the conventional pulsar
magnetosphere model contains a corotating, charge-separated plasma 
with number density 
\begin{eqnarray}
n_{GJ}({\bf r})=\frac{|{\bf \Omega}_B\cdot {\bf B}({\bf r})|}{2\pi e c(1-v^2/c^2)},
\label{eq:n_GJ_general}
\end{eqnarray} 
where $v$ is the corotation velocity (Goldreich \& Julian 1969).
We are primarily interested in the region of B's magnetosphere where 
$v\ll c$ and the magnetic field is approximately dipolar. For
our simplified model it suffices to ignore the angular dependence of the
field and set 
\begin{eqnarray}
&& B(r)\approx B_\star\left(\frac{R_\star}{r}\right)^3\, ,  \cr
\label{eq:B_dep}
&& {\bf \Omega}_B\cdot {\bf B}({\bf r})\approx \Omega_B B(r)\, ,
\label{eq:OMB}
\end{eqnarray}
where $R_\star$ is the neutron star radius and $B_0$ is its 
surface magnetic field.

The particle {\it number} density $n$ can be higher than $n_{GJ}$
because the addition of neutral plasma does not
affect the net charge density. We characterize this increase by the
parameter $\lambda\ge 1$ defined such that
\begin{eqnarray}
n(r)\approx \lambda(r)\frac{\Omega_B B(r)}{2\pi e c}\,  .
\label{eq:n}
\end{eqnarray} 
In what follows, we assume that the magnetospheric particles are
electrons and positrons.

Resonant cyclotron absorption is the dominant source of extinction for
radio waves passing through a conventional pulsar
magnetosphere. Cyclotron resonance occurs where 
\begin{eqnarray}
\omega\approx \omega_B\equiv \frac{eB}{m_ec},
\label{eq:omega_B}
\end{eqnarray}
with cross-section (Canuto, Lodenquai \& Ruderman 1971; 
Daugherty \& Ventura 1978) 
\begin{eqnarray}
\sigma(\omega)\approx \sigma_T\frac{\omega^2}
{(\omega-\omega_B)^2+\Gamma^2/4}\, .
\label{eq:cross_sect}
\end{eqnarray}
Here $\sigma_T$ is the Thompson cross-section and 
\begin{eqnarray}
\Gamma\equiv \frac{4e^2\omega_B^2}{3m_e c^3}
\label{eq:Gamma}
\end{eqnarray}
the natural line width. Strictly speaking, this cross section applies
to photons in particular modes which differ for electrons and positrons,
but we ignore this detail here. 

The peak cross section at resonance reaches $\sigma_{max}\approx 6\pi
(c/\omega_B)^2$, or roughly the square of the photon wavelength.
Nevertheless, the optical depth is modest because an incident photon
only resonates with cold electrons in a narrow radius range.  From
equations (\ref{eq:B_dep})-(\ref{eq:Gamma}) we find that resonance occurs at
radial distance from pulsar B of
\begin{eqnarray}
&& r_r(\omega)\approx R_\star\left(\frac{eB_\star}{m_e
c}\omega^{-1}\right)^{1/3}=
R_\star\left(\frac{\omega_{B\star}}{\omega}\right)^{1/3} \nonumber\\
&& \approx 1400~R_\star B_{\star 12}^{1/3}\nu_9^{-1/3}
\label{eq:r_r}
\end{eqnarray}
and that the optical depth is given by
\begin{eqnarray}
\tau_c\approx \frac{2\pi}{3}\lambda(r_r)\frac{\Omega_Br_r}{c}
\approx 0.2~\lambda(r_r)B_{\star 12}^{1/3}\nu_9^{-1/3}. 
\label{eq:opt_depth} 
\end{eqnarray}
Here $\nu_9$ is the radio frequency $\nu=\omega/(2\pi)$ expressed in GHz
and $\omega_{B\star}=1.8\times 10^{19}B_{\star,12}$ s$^{-1}$ is the 
cyclotron frequency at the surface of pulsar B (i.e. for $B=B_\star$)
with $B_{\star 12}\equiv B/(10^{12}~\mbox{G})$. Throughout the paper
we set the neutron star radius to be $10$ km and use $R_\star$ as a
unit of distance. Note that for the $\lambda=1$, $\tau_c$ is
approximately the ratio of the resonance radius $r_r$ to the radius of
B's light cylinder, $c/\Omega_B$, a result obtained previously by
Blandford \& Scharlemann (1976) and Mikhailovskii \etal (1982).

The determination of the inclination of the binary's orbit to the
plane of the sky based on the scintillation technique (Coles \etal
2004) implies that the radio beam of pulsar A passes pulsar B at an
impact parameter $p \sim 4\times 10^8$ cm. Since $p\lesssim r_r$, at a
first glance cyclotron absorption looks like a viable eclipse
mechanism, although a $\lambda\sim 10-100$ would be required to match
the eclipse depth. However, a closer look at cyclotron absorption, as
described below, reveals a problem: A's radio radiation heats the
particles in B's magnetosphere to relativistic energies making
synchrotron absorption rather than cyclotron absorption the relevant
process.

We denote by $F_\omega(\omega,r)$ the energy flux per unit frequency
from pulsar A at distance $r$ from pulsar B. A nonrelativistic
electron or positron absorbs and emits energy at rates (Rybicki \&
Lightman 1979)
\begin{eqnarray}
&& \dot E_+(r) = \frac{1}{2}\int
F_\omega(\omega,r)\sigma(\omega)d\omega \nonumber\\ &&
=2\pi^2\frac{e^2}{m_e c}F_\omega(\omega_B(r),r) 
\label{eq:E_dot_plus}
\end{eqnarray}
and 
\begin{eqnarray}
\dot E_-(r)=\frac{4}{9}\frac{e^4}{m_e^2 c^3}(\beta\gamma)^2B(r)^2.
\label{eq:E_dot_minus}
\end{eqnarray}
Here $\gamma\equiv (1-\beta^2)^{1/2}$ is the Lorentz factor of the
electrons and positrons which must be close to unity for cyclotron
absorption and emission to pertain. Balancing $\dot E_+$ by $\dot E_-$
yields
\begin{eqnarray}
\beta\gamma=\frac{3\pi}{2}\left[\frac{F_\omega(\omega_B(r),r)}
{\omega_B^2 m_e}\right]^{1/2}. 
\label{eq:beta_gamma}
\end{eqnarray}

Lyne \etal (2004) measure a time-averaged flux density $\approx 1.6$
mJy at $\nu_0\equiv\omega_0/(2\pi)=1.4$ GHz from pulsar A. Taking
$600$ pc for the distance to J0737-3039 one obtains $F_0\equiv
F_\omega(\omega_0) \approx 10^{-6}$ ergs cm$^{-2}$ at the position of
B which is separated from A by about $9\times 10^{10}$ cm. Using
equation (\ref{eq:beta_gamma}), we find that at $r_r\approx 1250 R_\star
B_{\star 12}$ where $1.4$ GHz photons get absorbed the electrons and
positrons have $\beta\gamma\approx 25$. This means that the electrons
and positrons are relativistic and that the cyclotron approximation is
inapplicable. The timescale for particles to become mildly
relativistic ($E\sim m_e c^2$) due to cyclotron absorption is
\begin{eqnarray}
t_{heat}^c(r)\approx\frac{m_ec^2}{E_+(r)}=
\frac{m_e^2 c^3/e^2}{2\pi^2F_\omega(\omega_B(r),r)},
\label{eq:t_cycl}
\end{eqnarray}
which is about $5$ s at the position where $1.4$ GHz photons
are resonantly absorbed. 

A related example of the heating of particles to relativistic energies
through resonant cyclotron absorption of radio waves is given in
Lyubarskii \& Petrova (1998). In their example the pulsar's radio
waves heat particles streaming along its open field lines.


\section{Synchrotron Absorption.}
\label{sect:synchro}


The mean cross-section for synchrotron absorption by an isotropic
distribution of particles with energy $\gamma m_e c^2$ is 
(Rybicki \& Lightman 1979)
\begin{eqnarray}
\sigma_s(\omega)\approx
\frac{8\pi^2}{3^{4/3}\Gamma(1/3)}\frac{e}{B}\left(
\frac{\omega_B}{\gamma \omega}\right)^{5/3}, 
\label{eq:sigma_synchro}
\end{eqnarray}
provided $\omega_B/\gamma\lesssim\omega\lesssim \gamma^2 \omega_B$.
Here $\Gamma(x)$ is a complete gamma function. 

In the absence of published measurements of pulsar A's radio spectrum,
we assume that it is a power law with most of the energy concentrated
at low frequencies (i.e. $F_\omega\propto\omega^{-\delta}$ with
$\delta>1$) as is typical for millisecond pulsars (Kuzmin \& Losovsky
2001).  The position of the low frequency cutoff of the spectrum is
not very important for our problem (although see \S
\ref{sect:reproc}).

For the cross-section (\ref{eq:sigma_synchro}) and an incident
spectrum with $\delta>1$, particle heating is dominated by the lowest
frequencies. This also implies that A's radio beam suffers a low
frequency cutoff which progresses toward higher frequency with
increasing depth in B's magnetosphere. Consequently, the absorbed
spectrum of A's radio flux at distance $r$ from pulsar B takes the
form
\begin{eqnarray}
F_\omega(\omega,r)=F_0\left(\frac{\omega_0}{\omega}\right)^{\delta}
\exp\left[-\tau_\omega(\omega,r)\right]\, ,
\label{eq:spectrum}
\end{eqnarray} 
where $\tau_\omega(\omega,r)$ is the frequency-dependent optical depth
at $r$. Using equations (\ref{eq:n}) \& (\ref{eq:sigma_synchro})  one finds
\begin{eqnarray}
&& \tau_\omega(\omega,r)=\int\limits_r^\infty
\sigma_s(\omega)n(r)dr\nonumber\\
&& \approx 
\frac{4\pi}{3^{4/3}\Gamma(1/3)\zeta_1}\lambda(r)\frac{\Omega_B r}{c}
\left(\frac{\omega_B}{\gamma \omega}\right)^{5/3},
\label{eq:tau_omega}
\end{eqnarray}
where in arriving at the last expression we have assumed that
$\lambda(r)(\omega_B/\gamma)^{5/3}$ is a steeply decreasing function 
of $r$ and 
\begin{eqnarray}
\zeta_1=4+\frac{5}{3}\frac{d\ln\gamma}{d\ln r}-\frac{d\ln\lambda}{d\ln r}.
\label{eq:zeta_1}
\end{eqnarray}
Expression (\ref{eq:tau_omega}) is valid provided $\omega\gtrsim
\omega_B/\gamma$; below this frequency the relativistic plasma becomes
transparent. Since extinction at observed frequencies occurs at $r\sim
10^9$ cm $\ll r_{L,B}=c/\Omega_B\approx 10^{10}$ cm, a large
$\lambda(r)$ is required to account for $\tau_\omega\gtrsim 1$.

Next we give a simplified evaluation of the energies to which
magnetospheric particles are heated assuming $\delta=2$ and
$\lambda$ independent of $r$.\footnote{Appendix \ref{ap:ap1} contains a more detailed
derivation.} We define a local cutoff frequency $\omega_1(r)$ such
that $\tau_\omega(\omega_1(r),r)=1$. Using equation (\ref{eq:tau_omega}) and
dropping constant coefficients we find
\begin{eqnarray}
\omega_1(r)\approx \frac{\omega_B(r)}{\gamma(r)}
\left[\lambda(r)\frac{\Omega_B r}{c}\right]^{3/5}.
\label{eq:omega_1}
\end{eqnarray}
Clearly $\omega_1(r)$ increases as $r$ decreases. Figure
\ref{fig:spectrum} illustrates how the low frequency part of
$F_\omega(\omega,r)$ erodes with increasing depth in B's
magnetosphere.\footnote{The low frequency cutoff of A's spectrum is
unimportant inside the radius at which the cutoff frequency falls
below $\omega_1(r)$.} The synchrotron heating rate, $\dot E_+^s$, is
is given by 
\begin{eqnarray}
&& \dot E_+^s(r)\approx \sigma(\omega_1(r))\omega_1(r) 
F_\omega(\omega_1(r),r) \nonumber\\
&& \approx F_0\left(\frac{\gamma \omega_0}{\omega_B}\right)^2
\frac{e}{B}\frac{\omega_B}{\gamma}\left[\lambda(r)\frac{\Omega_B r}
{c}\right]^{-8/5}.
\label{eq:q_synch_plus_raw}
\end{eqnarray}
which reflects the dominance of radiation with $\omega\sim\omega_1(r)$.  
Balancing the synchrotron heating rate by the synchrotron cooling rate  
(\ref{eq:E_dot_minus}), we obtain 
\begin{eqnarray}
&& \gamma(r)\approx 5\times 10^2\frac{F_0}{m_e \omega_{B\star}^2}
\left(\frac{\omega_0}{\omega_{B\star}}\right)^2\nonumber\\
&& \times\left[\lambda\frac{\Omega_B R_\star}
{c}\right]^{-8/5}
\left(\frac{r}{R_\star}\right)^{52/5}
\label{eq:gamma}\nonumber\\
&& \approx 2.5\times 10^4~ \lambda^{-8/5}
\left(\frac{r/R_\star}{10^3}\right)^{52/5}B_{\star 12}^{-4}\, .
\label{eq:gamma_2}
\end{eqnarray}
The calculation of $\gamma(r)$ tacitly assumes that synchrotron
absorption of A's radio emission is the only source for heating
particles in B's magnetosphere [cf. Lyutikov \& Thompson (2005)].

To evaluate the size of the eclipsing region $r_e(\omega)$, defined as
a distance from pulsar B at which $\tau_\omega(\omega,r_e)=1$, we
combine equations (\ref{eq:omega_1}) and (\ref{eq:gamma_2}) to arrive at
\begin{eqnarray}
&& r_e(\omega)=0.54~R_\star
\left(\frac{\omega_{B\star}}{\omega}\right)^{5/64}
\nonumber\\
&& \times\left[\frac{F_0}{m_e \omega_{B\star}^2}
\left(\frac{\omega_0}{\omega_{B\star}}\right)^2
\right]^{-5/64}
\left[\lambda\frac{\Omega_B R_\star}
{c}\right]^{11/64}\nonumber\\
&& \approx 380~R_\star \lambda^{11/64}\nu_9^{-5/64}
B_{\star 12}^{25/64}.
\label{eq:r_e}
\end{eqnarray}
The weak frequency dependence of $r_e$ is an artifact of the
assumed constancy of $\lambda$. 

The observed eclipse duration corresponds to $r_e\approx 10^9$ cm at
$\nu=1$ GHz. From equation (\ref{eq:r_e}) we find that this implies
$\lambda\approx 270$. For this value of $\lambda$, equation (\ref{eq:gamma_2})
yields $\gamma\approx 3.2~B_{\star 12}^{-4}$ at $r_e=10^9$ cm so our
assumption of synchrotron absorption is marginally valid.  The optical
depth $\tau_\omega$ cannot be arbitrarily high because radio photons
of frequency $\omega$ are not absorbed within the radius at which
$\omega\approx \omega_B/\gamma=\omega$. This ``saturation'' of
$\tau_\omega$ can be used to probe the dependence of $\lambda$ on $r$
(see equation [\ref{eq:ss_max_abs}] for a particular example).


\section{Enhancement of Particle Number Density in the 
Magnetosphere of Pulsar B.}
\label{sect:supply}


We hypothesize that there is a continuous supply of energetic
particles within the closed field line region of B's magnetosphere
from a source located close to the star where the magnetic field is
strong. Radiation damps the gyrational motions of particles born in
this region before they can slide out along magnetic field
lines. Under normal circumstances each particle would loop along the
closed field line on which it was born and strike the star's surface
in the opposite magnetic hemisphere. However, illumination by A's
radio beam can re-excite the gyrational motions of particles that stream
far away from B and then keep them suspended between
conjugate mirror points (see \S \ref{subsect:retention}). Thus, both
particle heating, making synchrotron absorption possible, and particle
trapping, leading to a higher optical depth eclipse, are mediated by
A's radio beam.


\subsection{Lifetimes of Energetic Particles.}
\label{subsect:lifetimes}

After illumination by A's radio beam ceases, particles trapped by
magnetic mirroring in B's magnetosphere cool and are lost. Synchrotron
emission reduces their energies to $m_e c^2$ on a timescale, $t_{cool}$
independent of their initial $\gamma\gg 1$. From
equation (\ref{eq:q_synch_plus1}), we estimate
\begin{eqnarray}
t_{cool}\sim \frac{m_e^3 c^5}{e^4 B^2}\approx 350~\mbox{s}~
\left(\frac{r/R_\star}{10^3}\right)^6B_{\star 12}^{-2} 
\label{eq:t_cool_synch}
\end{eqnarray}
As a consequence of relativistic beaming, components of momentum
parallel and perpendicular to ${\bf B}$ decay at the same rate.  In
the nonrelativistic regime, velocity components perpendicular to ${\bf
B}$ decay exponentially on timescale $t_{cool}$ while those parallel
to ${\bf B}$ remain nearly constant. As a result, the trapped
particles sediment onto the neutron star's surface on timescale
$t_{cool}$ leaving only the minimum number density of particles,
$n_{GJ}$, needed to support the corotation electric field.  Thus the
magnetosphere of pulsar B has to refill each time the radio 
beam of pulsar A illuminates it.


\subsection{Trapping of Particles by A's Beam.}
\label{subsect:retention}

The particle cooling time, $t_{cool}$, given by equation 
(\ref{eq:t_cool_synch}) is comparable to the local travel time,
$\sim r/(\beta_\parallel c)$, at the distance
\begin{eqnarray}
r_{cool}\approx 160R_\star \beta_\parallel^{-1/5}B_{\star 12}^{2/5}
\label{eq:r_cool}
\end{eqnarray} 
from pulsar B. Without heating, a particle arriving at $r_e$ from a
source located below $r_{cool}$ would be in its lowest gyrational
state. To become trapped, it has to gain enough perpendicular 
momentum during its outer magnetosphere passage 
to mirror above $r_{cool}$\footnote{The expression for the residence time,
$\sim r/(\beta_\parallel c)$, and that for $r_{cool}$ given by equation 
(\ref{eq:r_cool}) hold in both relativistic ($\beta_\parallel\approx 1$) 
and non-relativistic ($\beta_\parallel\ll 1$) regimes.} because of 
adiabatic invariant conservation.

The minimum gain in $p_\perp^2$ required for trapping is
\begin{eqnarray}
\Delta p_{\perp min}^2\gtrsim (\beta_\parallel\gamma)^2(m_e c)^2
\left(\frac{r_{cool}}{r_e}\right)^3,
\label{eq:p_perp_2}
\end{eqnarray}
where $\beta_\parallel$ is evaluated at $r_e$.  So the minimum energy,
$\Delta E_{min}$, that a particle has to absorb at $r\sim r_e$ is
\begin{eqnarray}
&& \Delta E_{min}\gtrsim \frac{\beta_\parallel^2\gamma}{2}m_e c^2
\left(\frac{r_{cool}}{r_e}\right)^3\nonumber\\
&& \approx 2\times 10^{-3} 
\beta_\parallel^{7/5}\gamma m_e c^2B_{\star 12}^{6/5}
\left(\frac{r_e/R_\star}{10^3}\right)^{-3}. 
\label{eq:E_min}
\end{eqnarray}

Cyclotron absorption of A's unattenuated radio 
beam in a single passage of $r_e$ results in an energy increase
\begin{eqnarray}
&& \Delta E_{+}\sim \int\limits^{r_e}\dot E_+(r)\frac{dr}{\beta_\parallel c}
\nonumber\\
&& \approx 2.3\times 10^{-4}\frac{m_e c^2}{\beta_\parallel}
\left(\frac{r_e/R_\star}{10^3}\right)^7, 
\label{eq:delta_E}
\end{eqnarray}
where the last line is evaluated for $\delta=2$.
Use of the cyclotron absorption formula (\ref{eq:E_dot_plus}) 
is appropriate because particles streaming past $r_e$ from a
source interior to $r_{cool}$ are at most mildly relativistic 
(see \S \ref{subsect:source}).  

Comparing $\Delta E_{min}$ with $\Delta E_{+}$ we find that 
trapping via cyclotron absorption and subsequent mirroring
is possible for particles with 
\begin{eqnarray}
\beta_\parallel^{12/5}\gamma\lesssim 0.12~B_{\star 12}^{-6/5}
\left(\frac{r_e/R_\star}{10^3}\right)^{10}.
\label{eq:cond}
\end{eqnarray}
Thus particles reaching $r_e\approx 10^9$ cm with $\beta_\parallel
\lesssim 0.4$ can be trapped.  This threshold is not very restrictive
(see \S \ref{subsect:source}), so the rate at which $\lambda$ can grow
is primarily determined by the rate at which particles are injected by
the source near B's surface.\footnote{The two-stream instability might
in principle assist in the trapping of particles, but an
unrealistically high number density is required for its growth time to
be comparable to the particle residence time in the magnetosphere.}


\subsection{Source of Particles.}
\label{subsect:source}

We can only speculate about possible sources of particles in the
corotating magnetosphere of pulsar B.  The requirement that B's
magnetosphere fill with an appropriate density of plasma while it is
illuminated by A's radio beam is not demanding. The total number of
particles needed to provide the corotating charge density in B's
magnetosphere is $N_{min}\sim (\Omega_B B_\star
R_\star^3/ec)\ln(r_{so}/R_\star)\sim 10^{30}$. Boosting this number by
a factor $\sim 10^2$ in $10^3$ s implies a trapping rate $\dot N\sim
10^{29}$ s$^{-1}$. If every trapped particle were born with the energy
$f_1 m_e c^2$ (subsequently lost as synchrotron radiation) and
only a fraction $f_2<1$ of them were trapped, the source power would
be $P\sim 10^{23}f_1/f_2$ ergs s$^{-1}$. For $f_1=10$ and $f_2=0.1$,
this amounts to $\sim 10^{25}$ ergs s$^{-1}$, much smaller than the
spin-down luminosity, $\sim 10^{30}$ ergs s$^{-1}$, estimated for
pulsar B by Lyne \etal (2004).

Our favored source is the creation of $e^\pm$ pairs on closed field
lines by gamma rays with energies $\sim 100$ MeV emitted by particles
accelerated in the outer gap of pulsar B (Cheng, Ho, \& Ruderman
1986a,b; Wang \etal 1998).\footnote{We are grateful to Jonathan Arons
for drawing our attention to this possibility.}  Photons from the
outer gap can enter the corotating magnetosphere and propagate at
significant angles to magnetic field lines. This facilitates $B\gamma$
pair creation close to the neutron star's surface.

Pairs born relativistic initially lose energy through synchrotron
radiation on a timescale $\Gamma^{-1}\sim 10^{-12}$ s $B_{10}^{-2}$
(see [\ref{eq:Gamma}]). As a consequence of relativistic beaming,
their pitch angles remain constant until they become
transrelativistic. Subsequent cooling by gyrosynchrotron radiation
completely damps gyrational motions but preserves the component of
velocity parallel to the magnetic field. Thus particles streaming away
from the neutron star have $\beta_\parallel\lesssim 1$. As we
discussed in \S \ref{subsect:retention}, this allows for their
efficient trapping when they are illuminated by A's radio beam.


\section{Steady State Magnetosphere.}
\label{sect:ss}

Under constant illumination by A's radio beam, B's magnetosphere would
achieve a steady state. We have already described how particle
energies would be set by a balance between the absorption and emission
of radiation. Here we concentrate on how the particle enhancement
factor $\lambda$ would be fixed in the context of the discussion in \S
\ref{sect:supply}

The efficiency of particle trapping drops sharply as the optical depth
at the local cyclotron frequency, $\tau_\omega(\omega_B(r),r)$,
increases above unity. The energy that fresh particles acquire by
cyclotron absorption in passing through the magnetosphere is reduced
by a factor $\exp[-\tau_\omega(\omega_B(r),r)]$ compared to $\Delta
E_+$ (see equation [\ref{eq:delta_E}]). Thus increasing
$\tau_\omega(\omega_B(r),r)$ lowers the maximum $\beta_\parallel$ at
which particle trapping can occur (see \S \ref{subsect:retention}).
The exponential dependence of trapping on optical depth suggests that
$\tau_\omega(\omega_B,r)$ does not greatly exceed unity and also that
it depends at most logarithmically on $r$.  We set
\begin{eqnarray}
\tau_\omega(\omega_B,r)=\chi\, ,
\label{eq:ss_abs_cond} 
\end{eqnarray}
where $\chi$ is a parameter we treat as independent of $r$ and whose
precise value depends upon the rate at which trapped particles are
lost by unspecified relaxation processes.

For $\omega>\omega_B/\gamma$ we find 
using equation (\ref{eq:tau_omega}) that $\tau_\omega(\omega,r)=\tau_\omega(\omega_B,r)
(\omega_B/\omega)^{5/3}$.  Thus, in the steady state magnetosphere
photons of frequency $\omega$ are absorbed at distance $r_e(\omega)$ 
such that $\omega_B(r_e)=\omega\chi^{-3/5}$. This gives 
\begin{eqnarray}
&& r_e(\omega)=\chi^{1/5}r_r(\omega)=R_\star\chi^{1/5}
\left(\frac{\omega_{B \star}}{\omega}\right)^{1/3}\nonumber\\
&& \approx 
1400~R_\star \chi^{1/5}B_{\star 12}^{1/3}\nu_9^{-1/3}
\label{eq:ss_r_e}
\end{eqnarray}
where $r_r(\omega)$ is defined in equation (\ref{eq:r_r}). In steady state 
the eclipse duration would scale as $\nu^{-1/3}$.

\begin{figure}
\plotone{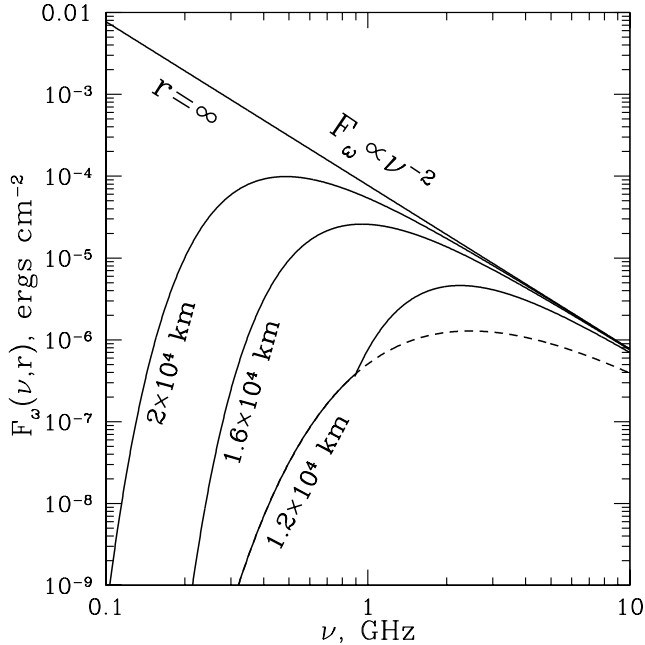}
\caption{
Evolution of pulsar A's radio spectrum due to 
synchrotron absorption in B's steady state magnetosphere assuming
$\chi=2$. Solid curves are labeled by distance from B. Dashed curve
illustrates maximally absorbed (with optical depth equal
to $\tau_{max}(\nu)$, see eq. [\ref{eq:ss_max_abs}]) 
spectrum close to B. 
\label{fig:spectrum}}
\end{figure}

The structure of the steady state magnetosphere for an arbitrary power
law spectrum of incident radio radiation is described by equations
(\ref{eq:ss_gamma}) and (\ref{eq:ss_lambda}) in Appendix
\ref{ap:ap2}. With typical parameters for J0737-3039 and $\delta=2$
we obtain
\begin{eqnarray}
&& \gamma(r)\approx 1.4\chi^{-24/55}
\left(\frac{r/R_\star}{10^3}\right)^{36/11}B_{\star 12}^{-12/11},
\label{eq:ss_gamma_2}\\
&& \lambda(r)\approx 10^2\chi^{3/11}
\left(\frac{r/R_\star}{10^3}\right)^{49/11}B_{\star 12}^{-20/11}.
\label{eq:ss_lambda_2}
\end{eqnarray}
Absorption by the extra plasma of A's radio radiation reduces the
efficiency of particle heating thus lowering the value of $\gamma$.
The dispersion measure variation during the eclipse caused by extra 
plasma within the steady state magnetosphere is at least 2
orders of magnitude below the current upper bound of $0.016$ pc
cm$^{-3}$ (Kaspi \etal 2004).

The important distinction of the steady state model 
is that it {\it predicts} $\lambda$. This determines
the lower and upper frequencies between which synchrotron
absorption is effective at a given $r$:
\begin{eqnarray}
&& \frac{\omega_B}{2\pi}\gamma^{-1}\approx 2~\mbox{GHz}~\chi^{24/55}
\left(\frac{r/R_\star}{10^3}\right)^{-69/11}B_{\star 12}^{23/11},\nonumber\\
&& \frac{\omega_B}{2\pi}\gamma^{2}\approx 5.6~\mbox{GHz}~\chi^{-48/55}
\left(\frac{r/R_\star}{10^3}\right)^{39/11}B_{\star 12}^{-13/11}\, .
\label{eq:ss_freqs}
\end{eqnarray} 
It follows that the maximum optical depth at frequency $\nu$, an
observable quantity, is given by
\begin{eqnarray}
\tau_{max}(\nu)\approx 3.2\chi^{15/23}\nu_9^{-20/23}\, .
\label{eq:ss_max_abs}
\end{eqnarray}
Figure \ref{fig:spectrum} depicts the evolution of the spectrum of A's
radio beam with depth in B's magnetosphere.  Power at low frequencies
is gradually eaten out by synchrotron absorption as the beam propagates
deeper into B's magnetosphere.

Illumination of B's magnetosphere by A's radio beam probably starts
only short time prior to eclipse. So we may be witnessing radio beam
attenuation by {\it dynamically evolving} plasma in B's
magnetosphere. The timescale for cold particles to become
trans-relativistic via cyclotron absorption evaluated from
equation (\ref{eq:t_cycl}) is rather short, typically $\lesssim 10$ s.
Energies of relativistic particles rise exponentially via synchrotron
absorption on timescale
\begin{eqnarray}
t_{heat}^s\approx 0.78\left[\frac{e^2}{m_e^2 c^3}F_\omega(\omega_B)\right]^{-1}
\approx
320~\mbox{s}~\left(\frac{r/R_\star}{10^3}\right)^{-6},
\label{eq:cool_synch}
\end{eqnarray}
so long as the radio spectrum of pulsar A remains unattenuated.
Heating has an exponential character for $\delta=2$ because as
$\gamma$ grows particles can absorb the incoming radio photons at
lower frequencies because $\omega_B/\gamma$ decreases making more
energy from the unabsorbed spectrum available to heat them. This
partly compensates for the decrease of the absorption cross section
with increase of $\gamma$ (see eq. [\ref{eq:sigma_synchro}]). Once
$\gamma$ grows so large that $\tau_\omega(\omega_B/\gamma,r)$ becomes
comparable to unity, heating is less efficient and $t_{heat}\propto
\gamma^{8/3}$. We estimate that particles reach $\gamma\sim 10^2$ at
$r\approx 10^9$ cm on a timescale of $\sim 5$ min. This is comparable
to estimates of several tens of minutes for the duration of the
illumination period judged from the orientation of pulsar A's spin and
dipole axes suggested by Demorest \etal (2004).  Not much can be said
about the timescale needed for the particle density to reach its
steady state value. Presumably this is largely controlled by the rate
at which the source near pulsar B is able to supply fresh trans-relativistic
particles since these are readily trapped.


\section{Reprocessed Radiation.}
\label{sect:reproc}

Energy absorbed by particles in B's magnetosphere from A's radio beam
is reemitted as synchrotron radiation, albeit with some time
delay. Absorption takes place at $\omega\sim \omega_1$, see
equation (\ref{eq:omega_1}), close to the minimum frequency possible,
$\omega_B/\gamma$, and is reemitted at the considerably higher
frequency $\omega\sim \gamma^2\omega_B$. This reemitted radiation,
although unimportant locally, may dominate the primary radiation from
pulsar A deeper in pulsar B's magnetosphere.

Here we attempt to calculate the properties of the reprocessed radiation
for the steady state magnetosphere by applying results from \S
\ref{sect:ss}. This requires choosing a low frequency cutoff,
$\nu_{min}$, for the assumed power law spectrum of A's radio
emission. Because there is little evidence for low-frequency cutoffs
in the spectra of millisecond pulsars above $100$ MHz (Kuzmin
\& Losovsky 2001), we normalize $\nu_{min}$ to this frequency. From
equation (\ref{eq:ss_r_e}) we deduce that the energy carried by photons with
$\nu\sim\nu_{min}$ is absorbed at a distance $r_{max}\approx 3500
R_\star\nu_{min,8}^{-1/3}$ from pulsar B and reemitted at frequency
$\nu_{max}\approx \gamma^2(r_{max})\omega_B(r_{max})/2\pi \approx
360\nu_{min,8}^{-13/11}$ GHz. The latter represents the upper cutoff
frequency of the reprocessed radiation because $\gamma^2\omega_B$
increases with increasing $r$ (see [\ref{eq:ss_gamma_2}]).  A detailed
calculation shows that radiation reemitted by all the relativistic
particles within $r_{max}$ has a flat power law spectrum $F_\nu$ with
index close to zero.  The local flux of reprocessed 
photons at $\nu_{max}$ can be estimated from $F_\nu(\nu_{max})\nu_{max}
\sim F_\nu(\nu_{min})\nu_{min}$ yielding 
$F_\nu(\nu_{max})\sim 2\times 10^{-9}$ ergs cm$^{-2}$ s$^{-1}$ Hz$^{-1}$ 
for a primary spectrum with $\delta=2$ and time-averaged intensity 
(at B's location) $F_\nu(1.4~\mbox{GHz})=6\times 10^{-6}$ 
ergs cm$^{-2}$ s$^{-1}$ Hz$^{-1}$. 
Reprocessed spectrum intersects the primary radio spectrum of
pulsar A at $\sim 50$ GHz and at frequencies higher than that 
one has to solve the radiation transfer problem to determine particle 
heating. Luckily, this is far enough from $1$ GHz region of spectrum
where eclipse observations are usually taken so that we do not need to
worry about such complications. Unfortunately, the reprocessed radiation 
can hardly be detected on Earth because of very small covering 
fraction of B's magnetosphere as seen from A.


\section{Discussion.}
\label{sect:disc}

A simple dynamical picture of pulsar A's eclipse by pulsar B's
magnetosphere emerges from our considerations.  Prior to the onset of
illumination by A's radio beam, the number density in the corotating
part of B's magnetosphere is equal to $n_{GJ}$. When A's radio beam
strikes the magnetosphere of pulsar B, perhaps $\sim 10$ min prior to
B's inferior conjunction, particles initially present in the
magnetosphere rapidly (in $\lesssim 10$ s) become relativistic. Their
energies continue to rise until either synchrotron absorption is
balanced by synchrotron emission or illumination by A's radio beam
ceases.  At the same time neutral plasma accumulates in the
magnetosphere as the result of the trapping of particles that stream
out in cold beams from a source near B. It is unclear whether the
magnetosphere reaches steady state prior to the eclipse or whether is
still evolving. In either case a density enhancement $\lambda\sim
200-300$ would be required to match the observed depth and duration of
the eclipse.  Some time, perhaps several tens of minutes, after the
eclipse illumination of B's magnetosphere by A's radio beam ceases and
the particles cool on a typical timescale $t_{cool}\sim 10^2-10^3$ s
given by equation (\ref{eq:t_cool_synch}). As a result particles no longer
mirror and all plasma beyond that needed to maintain the corotation
electric field is lost. This entire cycle repeats each orbital period.

We speculate that part of the asymmetry between eclipse ingress and egress
reflects a rise in plasma density in B's magnetosphere during the
$\sim 30$ s the eclipse lasts. The optical depth, $\tau_\omega$ is
very sensitive to increasing $\lambda$. Not only is $\tau_\omega$
directly proportional to the total column density of absorbing
particles, it is also proportional to the absorption cross section per
particle which varies as $\gamma^{-5/3}$ and $\gamma$ decreases with
increasing $\lambda$.  Thus if $\lambda$ were growing on a timescale
of minutes during eclipse, egress would be deeper than ingress and the
eclipse centroid would occur slightly after B reached inferior
conjunction since $r_e$ grows as $\lambda$ increases (see
eq. [\ref{eq:r_e}]). In this picture
the smooth ingress may partly be caused by the time-variable
optical depth at a fixed location rather than the $\tau_\omega$ variation
as the radio beam samples smaller values of $r$. On the contrary,
during egress the optical depth is higher and the abrupt termination of
eclipse may signal the emergence of the radio beam from
behind an almost opaque screen. This explanation requires proper timing.

Our model predicts a variation of the polarization of A's radio
emission during the course of the eclipse. The polarization signal
should be strongly correlated with B's rotational phase in a manner
similar to the eclipse lightcurve variations found by McLaughlin \etal
(2004b).  This prediction stems from the fact that synchrotron
absorption of photons in different polarization states is sensitive to
the angle between the photon ${\bf k}$ vector and the direction of the
magnetic field.  Modeling the polarization signature would be
facilitated because B's magnetic field should be nearly dipolar at
$r_e$ since $R_\star\ll r_e\ll r_{so}$.

Effects of general relativity are very important in J0737-3039.  Lai
\& Rafikov (2004) have demonstrated that gravitational light bending
can significantly (by $\sim 30\%$) change the minimum impact parameter
at which A's radio beam passes B. Thus gravitational lensing must play
a significant role in shaping the eclipse profile. Moreover, because
of the binary's finite orbital eccentricity, $e\approx 0.088$,
periastron precession driven mainly by the effects of general
relativity forces a 21 yr periodic variation, from $a|\cos i|(1-e)$ to
$a|\cos i|(1-e)$ neglecting lensing effects, of the pulsars' minimum
projected separation on the plane of the sky (Burgay \etal
2003). By changing the minimum impact parameter of A's radio beam
with respect to B, this should produce observable eclipse profile
variations.

Lyutikov (2004) and Arons \etal (2004) have suggested that the eclipse
of A's radio beam is caused by synchrotron absorption in the
magnetosheath that forms when A's relativistic wind impacts B's
magnetosphere. Detailed polarization observations offer a means to
distinguish this viable alternative from our model.

The model we have presented is necessarily rather simplistic --- it is
one-dimensional, it assumes that particle distribution functions are
isotropic, and so on. Where possible, future work should relax these
constraints. A quantitative description of the particle source giving
rise to the plasma density enhancement might also be pursued.  Further
observations have the opportunity to reveal additional clues to
properties in the eclipse region. Knowledge of A's radio spectrum is
crucial for calculating particle energies, trapping efficiencies, and
the eclipse duration. Time-resolved polarization of A's radio emission
during eclipse would constrain the magnetic field geometry and
particle distribution anisotropy. 

Clarifying details of this remarkable example of nonlinear coupling of
relativistic plasma to external radiation may provide clues for
understanding other puzzling phenomena such as the modulation of B's
radio emission by radiation from A (McLaughlin \etal 2004a) and the
periodic variations of the pulsar B's brightness that correlate with
its orbital phase (Ransom \etal 2004).

\acknowledgements 

We thank Jonathan Arons, Anatoly Spitkovsky and Sterl
Phinney for useful discussions and suggestions. We are especially 
grateful to Chris Thompson whose work on J0737-3039 parallels 
ours in many respects and who has graciously shared with us his 
results before publication. RRR acknowledges
support from the W. M. Keck Foundation and 
NSF grant PHY-0070928. Research by PMG was supported through NSF grant
AST-0098301.

\appendix


\section{Particle energy.}
\label{ap:ap1}

To properly compute the heating of particles by synchrotron 
absorption we evaluate
\begin{eqnarray}
\dot E_+^{s}(r)=\frac{1}{2}\int\limits_{\omega_B/\gamma}^\infty 
F_\omega(\omega,r)\sigma_s(\omega)d\omega\approx
\frac{4\pi^2}{3^{4/3}\Gamma(1/3)}\frac{F_0e}{B}\int\limits_0^\infty\left(
\frac{\omega_B}{\gamma \omega}\right)^{5/3}
\left(\frac{\omega_0}{\omega}\right)^{\delta}
e^{-\tau_\omega(\omega,r)}d\omega,
\label{eq:q_synch_plus}
\end{eqnarray}
where the factor $1/2$ roughly accounts for shadowing as B rotates.
The lower limit of integration is extended to zero (instead of
$\omega_B/\gamma$) because we are assuming that
$\tau_\omega(\omega_B/\gamma,r)\gg 1$.  The largest contribution to
the local heating comes from $\omega$ such that
$\tau_\omega(\omega,r)\sim 1$ which, coupled to the condition that
$\omega_B/\gamma<\omega$ and equation (\ref{eq:tau_omega}), assures
that $\tau_\omega(\omega_B/\gamma,r)\gtrsim 1$.

Substituting $\tau_\omega$ from equation (\ref{eq:tau_omega}) into 
equation (\ref{eq:q_synch_plus}) and calculating the integral over 
$d\omega$ we find
\begin{eqnarray}
\dot E_+^{s}(r)=
\frac{3\pi\zeta_1}{5}\Gamma\left(\frac{2+3\delta}{5}\right)
\left[\frac{4\pi}{3^{4/3}\zeta_1\Gamma(1/3)}\right]^{3(1-\delta)/5}
\frac{e\omega_B F_0}{B}\left(\frac{\omega_0}{\omega_B}\right)^{\delta}
\left[\lambda(r)\frac{\Omega_B r}{c}\right]^{-(2+3\delta)/5}
\gamma^{\delta-1}.
\label{eq:q_synch_plus1}
\end{eqnarray}
Balancing this heating rate by the cooling rate 
(\ref{eq:E_dot_minus}) yields
\begin{eqnarray}
\gamma(r)=\left[\frac{27\pi\zeta_1}{20}
\Gamma\left(\frac{2+3\delta}{5}\right)
\left(\frac{4\pi}{3^{4/3}\zeta_1\Gamma(1/3)}
\right)^{\frac{3(1-\delta)}{5}}\right]^{\frac{1}{3-\delta}}
\left[\frac{F_0}{m_e \omega_{B}^2}
\left(\frac{\omega_0}{\omega_{B}}\right)^\delta
\right]^{\frac{1}{3-\delta}}
\left[\lambda(r)\frac{\Omega_B r}
{c}\right]^{-\frac{2+3\delta}{5(3-\delta)}}.
\label{eq:gamma_full}
\end{eqnarray}
Substituting equation (\ref{eq:gamma_full}) into equation 
(\ref{eq:tau_omega}), assuming $\tau_\omega(\omega)=1$, and 
solving for $r_e$, we obtain the size of eclipsing region 
at frequency $\omega$:
\begin{eqnarray}
r_e(\omega)=R_\star
\left[\frac{4\pi}{3^{4/3}\Gamma(1/3)\zeta_1}\right]^{3/32}
\left[\frac{20}{27\pi\zeta_1\Gamma(\frac{2+3\delta}{5})}\right]^{5/64}
\left(\frac{\omega_{B\star}}{\omega}\right)^{\frac{5(3-\delta)}{64}}
\left[\frac{F_0}{m_e \omega_{B\star}^2}
\left(\frac{\omega_0}{\omega_{B\star}}\right)^\delta
\right]^{-5/64}
\left[\lambda\frac{\Omega_B R_\star}
{c}\right]^{11/64}.
\label{eq:r_e_full}
\end{eqnarray}
The constant coefficients in these expressions depend upon the values
of the parameters $\delta$ and $\zeta_1$.  For the radio
spectrum of pulsar A with $\delta=2$ and $\lambda$ independent of $r$, 
it follows that $\zeta_1=64/3$. Then 
the coefficient in equation (\ref{eq:gamma_full}) is $\approx 481$. 


\section{Steadily  illuminated magnetosphere.}
\label{ap:ap2}

Combining equations (\ref{eq:tau_omega}), (\ref{eq:ss_abs_cond}), and 
(\ref{eq:gamma_full}), $\gamma$ is self-consistently 
determined to be 
\begin{eqnarray}
\gamma(r)=\left[\frac{9\pi^2}{3^{1/3}5}
\frac{\Gamma\left(\frac{2+3\delta}{5}\right)}
{\Gamma(1/3)}\chi^{-\frac{2+3\delta}{5}}
\frac{F_0}{m_e\omega_B^2}\left(\frac{\omega_0}{\omega_B}
\right)^\delta\right]^{3/11}.
\label{eq:ss_gamma}
\end{eqnarray}
From equations (\ref{eq:tau_omega}), 
(\ref{eq:ss_abs_cond}), and (\ref{eq:ss_gamma}) we find that 
\begin{eqnarray}
\lambda(r)=\frac{3^{4/3}5\Gamma(1/3)}{4\pi}
\left[\frac{9\pi^2}{3^{1/3}5}
\frac{\Gamma\left(\frac{2+3\delta}{5}\right)}
{\Gamma(1/3)}\right]^{5/11}\chi^{\frac{3(3-\delta)}{11}}\frac{c}{\Omega_B r}
\left[\frac{F_0}{m_e\omega_B^2}\left(\frac{\omega_0}{\omega_B}
\right)^\delta\right]^{5/11}.
\label{eq:ss_lambda}
\end{eqnarray}
Equations (\ref{eq:ss_gamma}) and (\ref{eq:ss_lambda}) imply that 
$\zeta_1=5$, see equation (\ref{eq:zeta_1}). 
These formulae determine the structure of the steady state  magnetosphere for arbitrary $\delta>1$.


\end{document}